\documentclass[twocolumn,showpacs,preprintnumbers,amsmath,amssymb]{revtex4}   
\usepackage{graphicx}
\usepackage{dcolumn}
\usepackage{epsfig}
\usepackage{bbm}

\begin{document}

\newcommand{\beq}{\begin{eqnarray}}
\newcommand{\eeq}{\end{eqnarray}}
\newcommand{\Bsdll}{$B_{s,d}\to\mu^+\mu^-~$}
\newcommand{\Bqll}{$B_q\to\mu^+\mu^-~$}
\newcommand{\Bsmumu}{$B_s\to\mu^+\mu^-~$}
\newcommand{\Bdmumu}{$B_d\to\mu^+\mu^-~$}
\newcommand{\lae}{\stackrel{<}{\sim}}
\newcommand{\gae}{\stackrel{>}{\sim}}         
\newcommand{\no}{\nonumber\\}         
\newcommand\bfi{\begin{figure}}
\newcommand\efi{\end{figure}}

\title{ \Bsdll in  technicolor model with scalars }

\author{Zhaohua Xiong$^{1,2}$ and Jin Min Yang$^3$}
\affiliation{ $^1$ CCAST (World  Laboratory), P. O. Box 8730, Beijing 100080, China \\
              $^2$ Institute of High Energy Physics, Academia Sinica, Beijing 100039, China\\
              $^3$ Institute of Theoretical Physics, Academia Sinica, Beijing 100080, China}%
\date{\today}

\begin{abstract}
Rare decays \Bsdll are evaluated in technicolor model with 
scalars. $R_b$ is revisited to constrain the model parameter space.
It is found that restriction on $f/f'$ arising from $R_b$ 
which was not considered in previous studies requires $f/f'$ 
no larger than 1.9 at 95\% confidence level, implying 
no significantly enhancement for $Br(B_{s,d}\to\mu^+\mu^-)$ from 
neutral scalars in the model. However, the branching ratio of \Bsmumu  
can still be enhanced by a factor of 5 relative to the standard model 
prediction.  With the value of $f/f'\lae 1.9$, an upgraded Tevatron 
with an integrated luminosity $20fb^{-1}$ will be sensitive to enhancement
of \Bsmumu in this model provided that neutral scalar mass $m_\sigma$ 
is below 580 GeV. 
\end{abstract}

\pacs{12.60.Nz, 13.20.Hw, 13.38.Dg}

\maketitle

\section {Introduction} 
\label {sec:intro}

The flavor-changing neutral-current B-meson rare decays play
an important role for testing the Standard Model (SM) at loop level 
and probing new physics beyond the SM. Among these decays, \Bsdll are 
of special interest due to their relative cleanliness and good sensitivity 
to new physics.

There are numerous speculations on the possible forms of new physics, among
which supersymmetry and technicolor are the two typical different frameworks. 
Both frameworks are well motivated. As a low-energy effective theory, the 
technicolor model with scalars introduces additional scalars to
connect the technicolor condensate to the ordinary fermions \cite{sim-a}.  
The phenomenology of this model has been considered extensively in the literature 
\cite{xiong-a,sim-a,caronet,Chivukula:1992ap,sdk,evans,Carone:1995mx,yumian}.
It has been found that this model does not produce unacceptably large 
contributions to neutral meson mixings or to the electroweak $S$ and $T$ 
parameters\cite{sim-a,caronet}.  On the other hand, this  model does predict 
potentially visible contributions to b-physics observables such as $R_b$
\cite{Carone:1995mx} and the rate of various rare $B$-meson decays
\cite{Carone:1995mx, yumian, xiong-a}.

Studies \cite{Babu00} showed that the processes \Bsdll are sensitive to supersymmetry.
In this Letter we will extend our previous studies \cite{xiong-a,xiong}
and evaluate the branching ratio of \Bsdll in the technicolor 
model with scalars. First  we will present a brief description of the model,
then give the analytical calculations for \Bsdll.  We will focus our attention 
on the neutral scalars contributions, which are likely to be sizable because, as 
shown in our following analysis, they will be enhanced by a factor $(f/f')^4$ as 
the parameter $f/f'$ gets large.
Before performing the numerical calculations, we examine the current bounds 
on this model from a variety of experiments, especially the latest measurements of 
$R_b$\cite{rbvalue}. Since the theoretical expression for $R_b$ used in constraining 
the model parameter space \cite{Carone:1995mx} seems not right,
we will recalculate the contributions to $R_b$ from the scalars in this model. 
We find the constraint from $R_b$ is still strongest as indicated 
in \cite{sim-b}, compared with those from the direct searches for neutral and 
charged scalars \cite{pdg}, $B^0-\bar{B^0}$ mixing, $b\to s\gamma$ \cite{bsgamma} 
as well as the muon anomalous magnetic moment \cite{E821}. Further, 
we evaluate restriction on $f/f'$ arising from $R_b$ which was not considered
in previous studies. Subject to the 
current bounds, the numerical results are presented in Sec.~\ref{result}.
Finally, the conclusion is assigned in Sec.~\ref{conclusion}.

\section{The technicolor model with scalars}
\label{model}

In this section we will briefly discuss the technicolor model with scalars 
and give the relevant 
Lagrangians which are needed in our calculations. More details of the model 
have been described in Refs.~\cite{sim-a, caronet}.

The model embraces the full SM gauge structure and 
all SM fermions which are technicolor singlets. It has a minimal $SU(N)$ 
technicolor sector, with two techniflavors that transform as a left-handed doublet 
and two right-handed singlets under $SU(2)_W$, 
\begin{eqnarray}
T_L=\left(
\begin{array}{c}
p \\ 
m
\end{array} \right)_L,  \,\,\,\,\, p_R, \,\,m_R
\end{eqnarray}
with weak hypercharges $Y(T _L)=0$, $Y(p_R)=1$, and $Y(m_R)=-1$.  
All of the fermions couple to a weak scalar
doublet $\phi$ to which both the ordinary fermions and technifermions are coupled.
This scalar's purpose is to couple the technifermion condensate to the
ordinary fermions and thereby generate fermion masses.  
If we write the matrix form of the scalar doublet as
\begin{equation}
\Phi=\left[ \begin{array}{cc}
            \bar{\phi}^0 & \phi^+\\
            -\phi^-      & \phi^0
            \end{array} \right ]
\equiv \frac{(\sigma+f')}{\sqrt{2}}\Sigma',
\end{equation}
and adopt the non-linear representation $\Sigma=exp(\frac{2i\Pi}{f})$
and $\Sigma'=exp(\frac{2i\Pi'}{f'})$ for technipion,
with fields in $\Pi$ and $\Pi'$ representing the pseudoscalar bound 
states of the technifermions $p$ and $m$,
then the kinetic terms for the scalar fields are  given by
\beq
{\cal L}_{K.E.}&=&\frac{1}{2}\partial_\mu\sigma\partial^\mu\sigma
+\frac{1}{4}f^2Tr({D}_\mu\Sigma^\dagger {D}^\mu\Sigma)\no
&&+\frac{1}{4}(\sigma+f')^2Tr(D_\mu{\Sigma'}^\dagger D^\mu\Sigma').
\label{kinetic}
\eeq  
Here $D^\mu\ ({D'}^\mu)$ denote the $SU(2)_L\times SU(2)_R$ covariant derivatives, 
$\sigma$ is an isosinglet scalar field, $f$ and $f{'}$ are the 
technipion decay constant and the effective vacuum expectation value (VEV), 
respectively. 

As  mixing between $\Pi$ and $\Pi^{'}$ occurs, 
$\pi_a$ and $\pi_p$ are formed with $\pi_a$ becoming 
the longitudinal component of the W and Z, 
and  $\pi_p$ remaining in the low-energy theory as an isotriplet of physical 
scalars. From Eq.\ (\ref{kinetic}) one can obtain the correct gauge boson masses
providing that $f^2+f^{'2}=v^2$ with the electroweak scale $v=246\ GeV$.

Additionally, the contributions to scalar potential generated by 
the technicolor interactions should be included in this model. The
simplest term one can construct is
\begin{equation}
{\cal L}_T=c_14\pi f^3Tr\left[\Phi\left(
\begin{array}{cc}
h_+ & 0\\
0 & h_-
\end{array}
\right)
\Sigma^\dagger\right] +h.c.,
\label{poential}
\end{equation}
where $c_1$ is a coefficient of order unity, $h_+$ and $h_-$ are the 
Yukawa couplings of scalars to $p$ and $m$ . 
From Eq.\ (\ref{poential}) the mass of the 
charged scalar at lowest order is obtained as  
\beq
m_{\pi_p}^2=2c_1\sqrt{2}\frac{4\pi f}{f'} v^2 h
\label{mpip}
\eeq
with $h\equiv (h_+ + h_-)/2$. To absorb the largest 
Coleman-Weinberg  radiative corrections ~\cite{Coleman:Weinberg}
for the $\sigma$ field which affect the phenomenology of the
charged scalar, the shifted scalar mass $\widetilde{M}_\phi$ and
coupling $\tilde{\lambda}$ are determined by
\beq
{\widetilde{M}_\phi}^2f'+\frac{\tilde{\lambda}}{2}{f^\prime}^3
=8\sqrt{2}c_1\pi hf^3.
\eeq
Therefore, the mass of the scalar $\sigma$ can be  expressed as  
\beq
\label{shifted:mass}
m_{\sigma}^2=\widetilde{M}_\phi^2
+\frac{2}{3\pi^2}\left[6\left(\frac{m_t}{f'}\right)^4+Nh^4\right]{f'}^2
\eeq
in limit ($i$) where the shifted $\phi^4$ coupling $\tilde{\lambda}$ is 
small and can be neglected and 
\beq
m_{\sigma}^2=\frac{3}{2}\tilde{\lambda}{f^\prime}^2 
    - \frac{1}{4\pi^2}\left[6\left(\frac{m_t}{f'}\right)^4+Nh^4\right]{f'}^2.
\eeq
in limit ($ii$) where the shifted mass of the scalar doublet $\phi$, 
$\widetilde{M}_\phi$ is small and can be neglected. 
The advantage of this model is at the lowest 
order, only two independent parameters in the limits ($i$) and 
($ii$) are needed to describe the phenomenology.  
We choose $(h,m_\sigma)$  as physical parameters and assume  
$N=4$ and $c_1=1$ in numerical calculations.

\section{Calculations}
\label{calculation}

We start the calculation by writing down the effective Hamiltonian
describing  the process \Bqll (q=s,d)
\beq
{\cal M}&=&\frac{\alpha G_F}{\sqrt{2}\pi}V_{tb}V_{tq}^*
\left\{-2C_7^{eff}\frac{m_b}{p^2}~\bar{q}i\sigma_{\mu\nu}p^\nu P_Rb\right.\no
&&\left.+C_9^{eff}~\bar{q}\gamma_\mu P_Lb ~\bar{\mu}\gamma^\mu\mu
+C_{10}~\bar{q}\gamma_\mu P_L b ~\bar{\mu}\gamma^\mu\gamma_5\mu
\right.\no
&&\left.+C_{Q_1}~\bar{q}P_Rb ~\bar{\mu}\mu
+C_{Q_2}~\bar{q}P_Rb ~\bar{\mu}\gamma_5\mu\right\} ,
\label{matrix}
\eeq
where $P_{R,L}=\frac{1}{2}(1\pm\gamma_5)$, $p$ is the momentum transfer. 
Operators ${\cal O}_{7,9,10}$ which correspond to the first three Wilson 
coefficients are the same as those given in \cite{Grinstein89} and ${\cal Q}_{1,2}$
corresponding to the last two are the additional operators arising from the 
neutral scalars exchange diagrams~\cite{Dai97} .

Using the effective Hamiltonian and 
\beq
\langle 0|\bar{q}\gamma_\mu\gamma_5 b|B_q\rangle=-f_{B_q}p_\mu,\no
\langle 0|\bar{q}\gamma_5 b|B_q\rangle=-f_{B_q}m_{B_q},\no
\langle 0|\bar{q}\sigma_{\mu\nu}(1+\gamma_5) b|B_q\rangle=0,
\label{gmb3} 
\eeq
we find that only operator ${\cal O}_{10}$ and ${\cal Q}_{1,2}$ 
contribute to process \Bqll with the decay rate given by
\beq
& & \Gamma (B_q\to\mu^+\mu^-)=\frac{\alpha^2G_F^2}
{64\pi^3}\left\vert V_{tb}V_{tq}^*\right\vert^2f_{B_q}^2m_{B_q}^3\no
& & ~~~~~~~~~~~~\times\left[C_{Q_1}^2
+\left(C_{Q_2}+\frac{2m_\mu}{m_{B_q}}C_{10}\right)^2\right].
\eeq
For convenience, we write down the branching fractions numerically
\beq
Br(B_d\to\mu^+\mu^-)&=&3.8\times 10^{-9}
\left[\frac{\tau_{B_d}}{1.65ps}\right]
\left[\frac{f_{B_d}}{210 MeV}\right]^2\no
&&\times\left\vert \frac{V_{td}}{0.008}\right\vert^2
\left[\frac{m_{B_d}}{5.28 GeV}\right]^3\left[C_{Q_1}^2\right.\no
&&\left.+\left(C_{Q_2}+2\frac{m_\mu}{m_{B_d}}C_{10}\right)^2\right],\no
Br(B_s\to\mu^+\mu^-)&=&1.2\times 10^{-7}
\left[\frac{\tau_{B_s}}{1.49ps}\right]
\left[\frac{f_{B_s}}{245 MeV}\right]^2\no
&&\times\left\vert \frac{V_{ts}}{0.04}\right\vert^2
\left[\frac{m_{B_s}}{5.37GeV}\right]^3\left[C_{Q_1}^2\right.\no
&&\left.+\left(C_{Q_2}+2\frac{m_\mu}{m_{B_s}}C_{10}\right)^2\right],
\label{decaywidth}
\eeq
where $\tau_{B_q}$ and $f_{B_q}$ are the $B_q$ lifetime and 
decay constant, respectively.
 
In the technicolor model with scalars, the additional contributions arise from 
the scalars. The contributions of the charged scalar $\pi_p^{\pm}$
with gauge boson $Z,\gamma$ exchanges to the Wilson coefficients $C_{10}$ 
at $m_W$ scale have been calculated by using Feynman rules 
derived from Eq.~(\ref{kinetic}),~(\ref{poential}) and given 
by~\cite{yumian,xiong-a}
\beq
{C}_{10}(m_W)_{TC}&=&\frac{x_W}{\sin^2\theta_W}(\frac{f}{f^{'}})^2
\left[-\frac{x_{\pi_p}}{8(x_{\pi_p}-1)} \right.\no
&&\left.+\frac{x_{\pi_p}}{8(x_\pi-1)^2}\ln\ x_{\pi_p}\right] ,
\eeq
where $\theta_W$ is the Weinberg angle and $x_i=m_t^2/m_i^2$. 
As for the contributions arising from the neutral scalars exchanges, when 
only the leading terms in large $f/f^{'}$ limit kept, they can be 
expressed as
\cite{xiong-a} 
\beq
C_{Q_1}(m_W)_{TC}&=&-\frac{x_W}{\sin^2\theta_W}(\frac{f}{f^{'}})^4
\frac{m_bm_\mu}{m_\sigma^2}\left[\frac{4x_{\pi_p}^2-7x_{\pi_p}+1}
{16(x_{\pi_p}-1)^2}\right.\no
&&\left.-\frac{x_{\pi_p}^2-2x_{\pi_p}}{8(x_{\pi_p}-1)^3}
\ln\ x_{\pi_p}\right] , \no
C_{Q_2}(m_W)_{TC}&=&-\frac{x_W}{\sin^2\theta_W}(\frac{f}{f^{'}})^4
\frac{m_bm_\mu}{m_{\pi_p}^2}\left[\frac{x_{\pi_p}+1}{8(x_{\pi_p}-1)}
\right.\no
&&\left.-\frac{x_{\pi_p}}{4(x_{\pi_p}-1)^2}\ln\ x_{\pi_p}\right].
\label{cq12w}
\eeq

From Eqs. (\ref{decaywidth}-\ref{cq12w}) we find that 
(1) both the contributions arising from the neutral scalar exchange 
$C_{Q_{1,2}}$ and gauge boson exchange $C_{10}$ are subject to helicity 
suppression,
(2) the contributions arising from the neutral scalar exchanges are proportional to 
$(f/f^{'})^4$, while those from the gauge bosons exchanges proportional 
to $(f/f')^2$. 
So for a sufficiently large $f/f'$, the contributions of neutral scalar exchanges
 are relatively enhanced and may become comparable with those from the
gauge boson exchanges.

The Wilson coefficients at the lower scale of 
about $m_b$ can be evaluated down from $m_W$ scale by using the 
renormalization group equation. At leading order, the Wilson coefficients 
are
\cite{Grinstein89,Dai97}
\begin{eqnarray} 
C_{10}(m_b)&=&C_{10}(m_W),\\
C_{Q_i}(m_b)&=&\eta^{-\gamma_Q/\beta_0}C_{Q_i}(m_W).
\end{eqnarray}               
where $\beta_0=11-2n_f/3$, $\eta=\alpha_s(m_b)/\alpha_s(m_W)$ and $\gamma_Q=-4$ 
is the anomalous dimension of $\bar{q} P_R b$.   

\section{Constraints from $R_b$}
\label{constraints}

Before presenting the numerical results,  let us  consider 
the current bounds on technicolor with scalars
from a variety of experiments, especially the measurement of $R_b$.  
Using the Feynman rules in Ref.~\cite{xiong-a}, 
one can easily find that the contributions from neutral scalars are negligible 
compared with those from charged scalars which appear in Fig.~\ref{Fig:feyman}, 
and the bottom mass-dependent terms in $R_b$ can also be omitted safely. 
In these approximations the addition contribution in the technicolor 
with scalars is obtained as
\beq
\delta R_b=R_b^{SM}(1-R_b^{SM})\Delta^{TC}
\eeq
with 
\beq
\Delta^{TC}&=&(\frac{f}{f'})^2\frac{\alpha}{4\pi\sin^2\theta_W} 
\frac{m_t^2}{m_W^2}\frac{v_{bL}}{v_{bL}^2+v_{bR}^2}
\left\{v_{bL}B_1\right.\no
&&\left.+v_{tR}\left[
m_Z^2(C^a_{22}-C^a_{23})+2C^a_{24}-\frac{1}{2}\right]\right.\no
&&\left.-2v_{tL}m_t^2C^a_0-\cos2\theta_WC^b_{24}\right\}.
\label{rbtc}
\eeq
\bfi
\epsfig{file=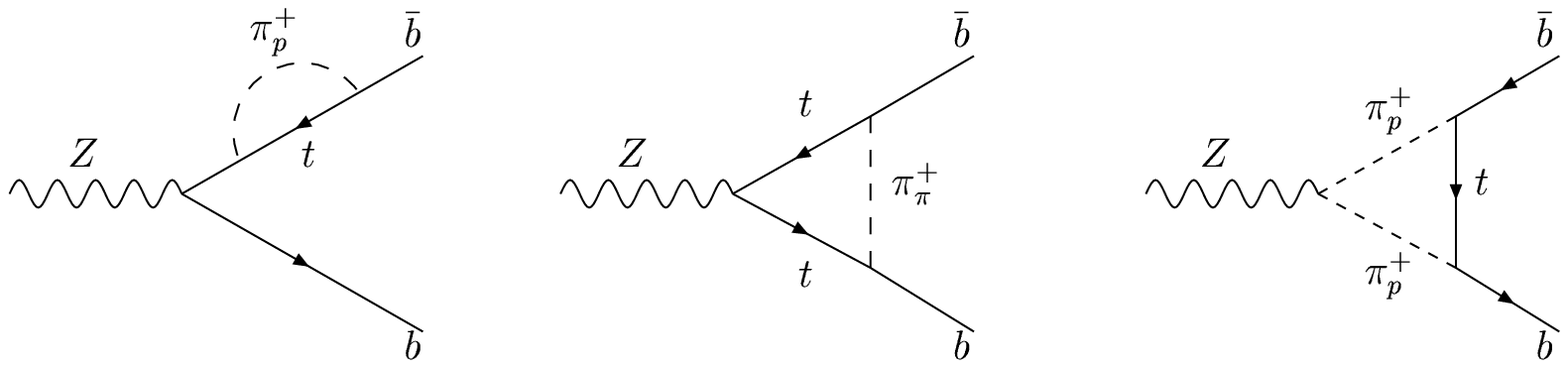,width=200pt,height=100pt} 
\caption{Charged scalars diagrams contributing to $Zb\bar{b}$.}
\label{Fig:feyman}
\efi
Here $B_1=B_1$($-p_1$,$m_t$,$m_{\pi_p}$), 
$C^a_{0,ij}=C_{0,ij}$($p_1$,$-P$, $m_{\pi_p}$, $m_t$, $m_t$) and $C^b_{24}
=C_{24}(-p_1,P,m_t,m_{\pi_p},m_{\pi_p})$, with  $p_1(p_2)$ and $P$ 
denoting the four-momentum of  $b (\bar{b})$ and Z boson
respectively, are the Feynman loop integral functions and their expressions
can be found in \cite{bcanaly}. The coupling constants $v_{qL}$ and $v_{qR}$  
are given by
\beq
v_{qL}=T^q_3-e_q\sin^2\theta_W,~~ v_{qR}=-e_q\sin^2\theta_W.
\eeq
Our explicit expressions are not in agreement with those used in
\cite{Carone:1995mx} where the results obtained in the framework of 
the two-Higgs doublet model (THMD) \cite{Boulware91} were adopted directly. 
We checked the calculations and confirmed our results. 

\bfi[hbt]
\epsfig{file=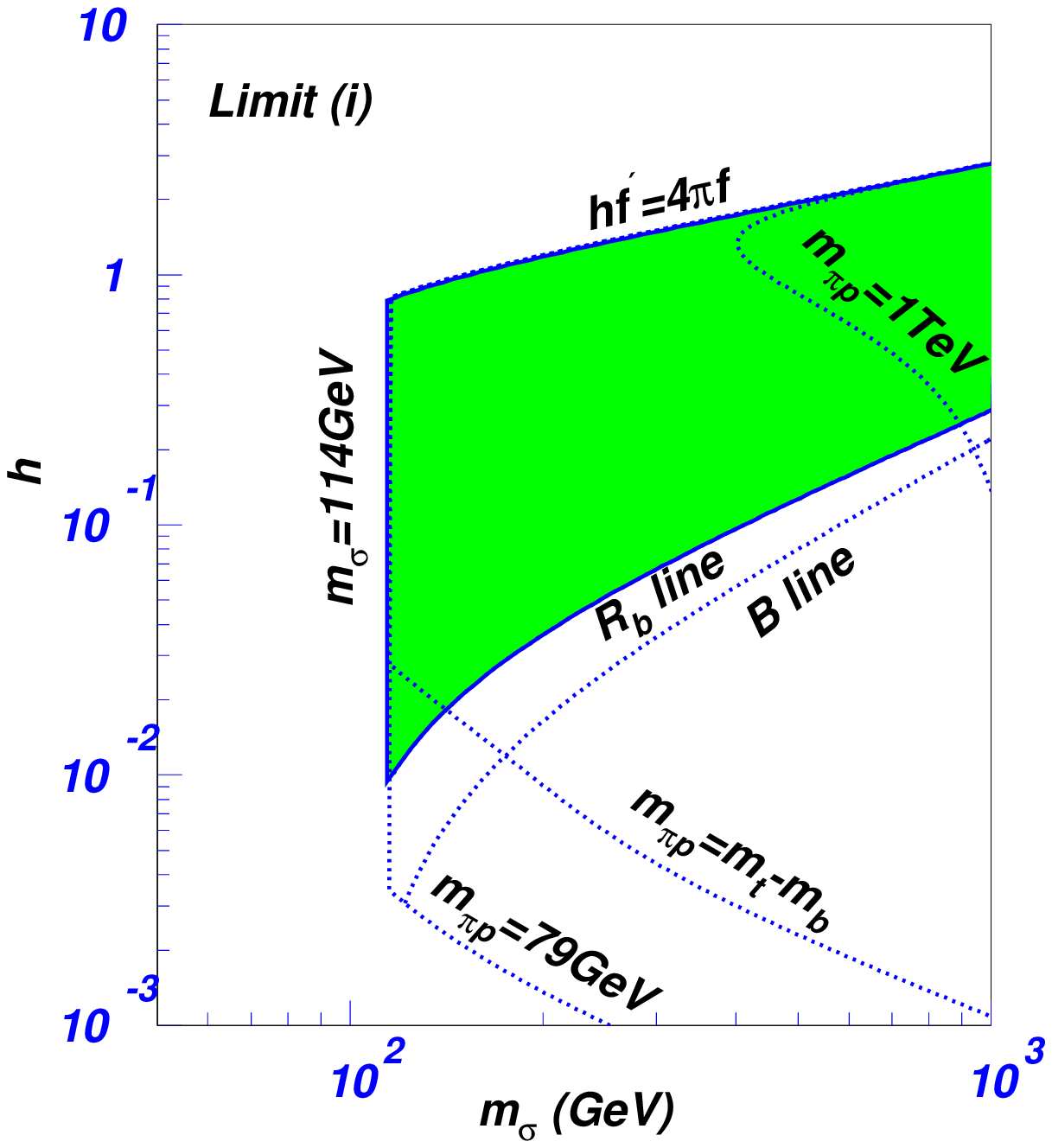,width=8cm} 
\caption{\small Constraints on technicolor with scalars in limit $(i)$.
The allowed parameter space is the shaded region bounded by the contours 
$m_\sigma = 114$ GeV, $\delta R_b$ ($R_b$ line) and $h f' = 4 \pi f$.  
The current bound from the searches for charged scalars 
$m_{\pi_p} =79$ GeV is shown along with the reference curves
$m_{\pi_p} = m_t - m_b$, $m_{\pi_p} = 1\ TeV$. 
The constraint from $B^0-\bar{B}^0$ mixing is labeled ``B-line''.} 
\label{Fig:Rb1}
\efi

The current measurement of $R_b$ reported by the LEP is 
$R_b^{expt} = 0.21646 \pm 0.00065$ \cite{rbvalue}.
Comparing with the SM value $R_b^{SM} =0.21573\pm 0.0002$,
we obtained the constraints in $h$ versus $m_{\sigma}$ plane
shown in Figs.~\ref{Fig:Rb1} and ~\ref{Fig:Rb2}. 
Although our explicit expression for $R_b$ is different from 
that used in \cite{sim-b}, a comparison of Fig.~\ref{Fig:Rb1}, 
Fig.~\ref{Fig:Rb2} with Figure 1 in Ref.~\cite{sim-b}
suggests that there is not a qualitative change in the results plotted.

Our numerical results show that the constraint on $f/f'$ from $R_b$ is quite 
stringent, i.e., the ratio of $f/f'$ must be smaller than 1.9 at 95\% C. L., implying 
that the neutral scalars will not give dominate contributions to 
the processes of \Bsdll. 
Since previous studies did not comment on any restriction 
on $f/f'$ arising from $R_b$, this is a new and interesting conclusion.

\bfi[hbt]
\epsfig{file=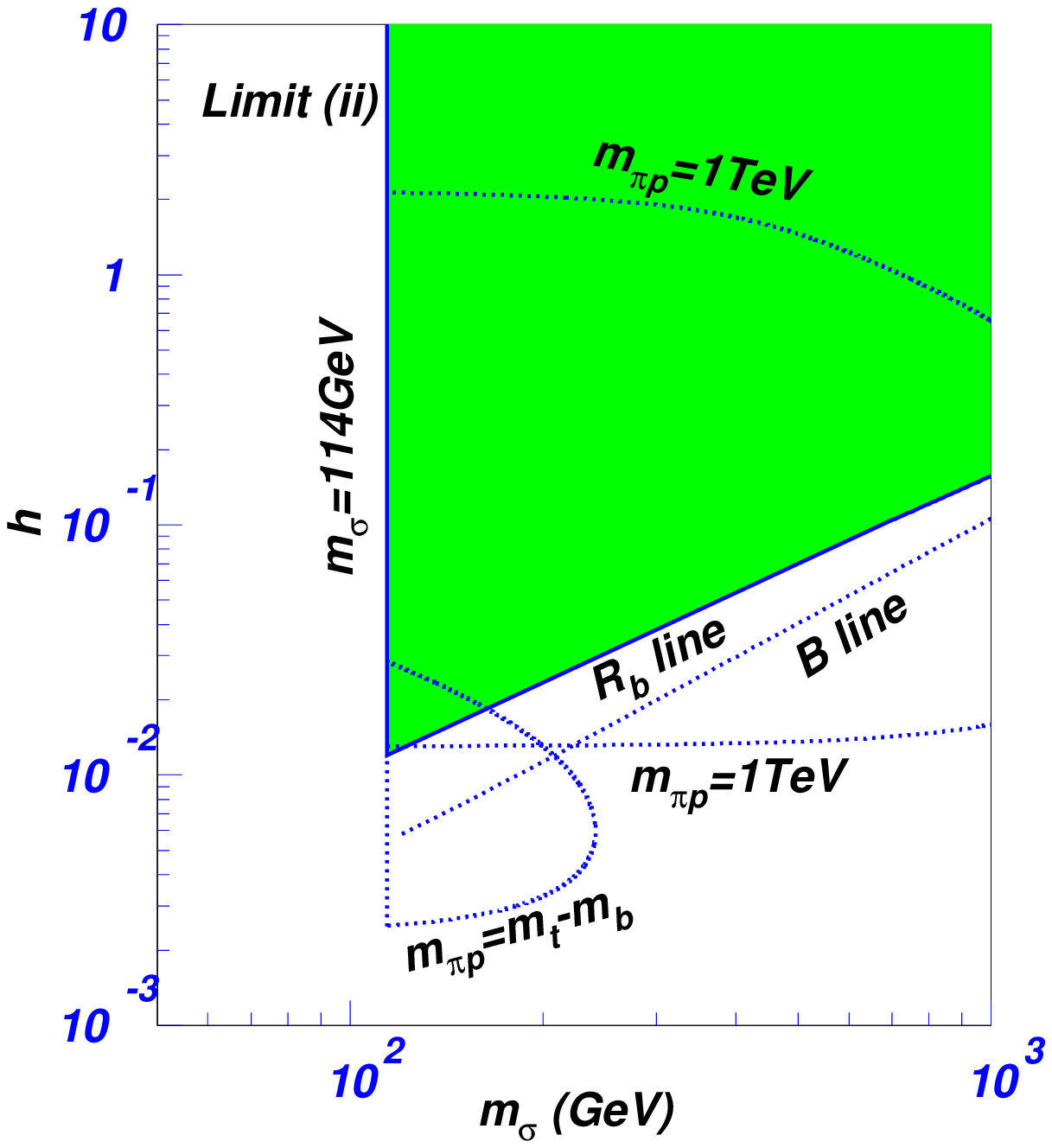,width=8cm} 
\caption{\small Constraints on technicolor with scalars in limit $(ii)$.
The allowed parameter space is the shaded region bounded by the contours 
$m_\sigma = 114$ GeV and $\delta R_b$ ($R_b$ line).
Other bound curves are the same as Fig.~\ref{Fig:Rb1}.} 
\label{Fig:Rb2}
\efi

In Figs.~\ref{Fig:Rb1} and \ref{Fig:Rb2} we also display the bounds from 
$B^0-\bar{B}^0$ mixing and from the limits of Higgs masses \cite{sim-b}.
In technicolor theories where the charged scalars 
couple to fermions in a similar pattern as in type-I two-Higgs 
doublet model, the strongest limit $m_{\pi^\pm_p}\geq 79$ GeV
has been obtained directly from LEP experiments~\cite{pdg}.
On the other hand, the LEP collaborations \cite{pdg}
have placed a 95\% C. L. lower limit on the SM Higgs boson
$M_H^0\geq 113.5$ GeV from searching for the process $e^+e^-\to Z^* \to Z H^0$. 
Although the limit on technicolor scalars may differ from that 
on $M_H^0$, in practice, the contour $m_\sigma=114$ GeV can serve as 
an approximate boundary to the experimentally allowed region \cite{caronet,sim-b}. 
Note that the chiral Lagrangian analysis break down  only 
constrain on the parameter space in limit $(i)$ \cite{Carone:1995mx}, the area 
above and to left of $hf'=4\pi f$ line is excluded because the technifermion 
current masses are no longer small compared to the chiral symmetry breaking 
scale. For references, we also plotted the contours $m_{\pi_p}=m_t-m_b$ and 
$m_{\pi_p}=1\ TeV$. If the top quark doesn't decay to $\pi_p^+b$, 
the areas {\em outside} of $m_{\pi_p}=m_t-m_b$ curve is excluded in 
Fig.~\ref{Fig:Rb1}.
Similar situation occurs to $m_{\pi_p}=1\ TeV$ curve in Fig.~\ref{Fig:Rb2}
if all scalar masses are restricted to the sub-TeV regime.
In contrast to these, the excluded parameter space 
are the areas {\em inside} of $m_\pi=m_t-m_b$ curve
in limit ($ii$) and $m_{\pi_p}=1\ TeV$ curve in limit ($i$). 

The constraint from $b\to s \gamma$ is close to that from $B^0-\bar{B}^0$ 
mixing~\cite{xiong-a,yumian,gsw,mbsg}, which are weaker than those from 
$R_b$~\cite{Carone:1995mx}.  As for the constraints 
from the measurement of $g_\mu-2$, our previous study \cite{xiong-b} 
showed that if the deviation of the E821 experiment result \cite{E821} 
and SM prediction 
$\Delta a_\mu\equiv a_\mu^{exp}-a_\mu^{SM}=(43\pm 16)\ 10^{-10}$ 
persists,  it would severely constrain the technicolor models because the
technicolor models can hardly provide such a large contribution. 
However, over the last year the theoretical prediction of $a_\mu$ in the SM 
has undergone a significant revision due to the change in sign of the light 
hadronic correction, which leads to only a 1.6$\sigma$
deviation from the SM \cite{Knecht01}, yielding no more useful limits
on this model.

\section{Numerical results}
\label{result}

Bearing the constraints on technicolor with scalars in mind, and 
for the same values of $m_{\pi_p}$ and $f/f'$, the allowed value of 
$m_{\sigma}$ is generally smaller in limit $(i)$, from 
Eq.(\ref{cq12w}) one can infer  easily that 
the additional contributions to \Bsdll in limit $(i)$ will be larger than 
those in  limit $(ii)$.  Furthermore, as can be seen from the numerical 
coefficients in Eq.~(\ref{decaywidth}), the decay rate of $B_s$ is significantly 
larger than $B_d$ due  primarily to the relative size of $|V_{ts}|$ to $|V_{td}|$.  
We thus take the $B_s$ decay in limit $(i)$ as 
an example to show the numerical results.

\bfi[hbt]
\epsfig{file=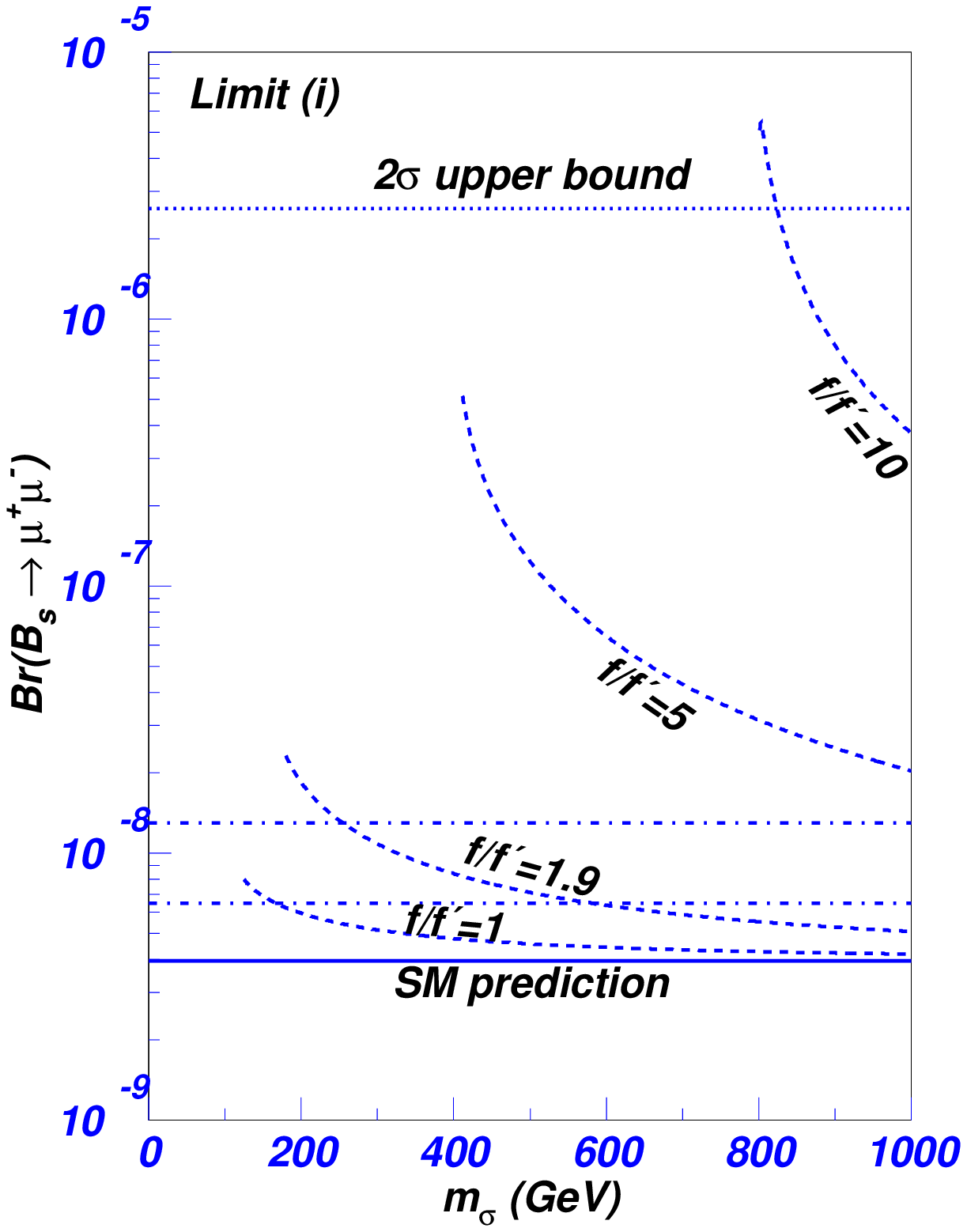,width=8cm} 
\caption{\small $Br(B_s\to\mu^+\mu^-)$ as a function of $m_\sigma$
for $f/f'=1,1.9,5,10$ (dash lines) in limit $(i)$. 
The current $2\sigma$ upper bound \cite{CDF} (dotted line), 
the SM prediction (solid line) as well as the expected 
sensitivity of the upgraded Tevatron with $10fb^{-1}$ and $20fb^{-1}$ 
(the dash-dotted lines) are also shown.}
\label{Fig:bsmumu}
\efi
The experimental bound on \Bsmumu comes from the CDF\cite{CDF} 
\beq 
Br(B_s\to\mu^+\mu^-)<2.6\times 10^{-6}
\eeq
at 95\% C.L. with the corresponding integrated luminosity about $100pb^{-1}$,
while the  SM prediction
\beq
Br(B_s\to\mu^+\mu^-)= 4.0\times 10^{-9}   
\eeq
is obtained by taking the central values for all inputs in 
Eq.~(\ref{decaywidth}). 
The branching ratio of \Bsmumu as a function of $m_\sigma$ is displayed   
in Fig.~\ref{Fig:bsmumu} for various values of $f/f'$. 
The $2\sigma$ bounds at the upgraded Tevatron with  
$10fb^{-1}$ and $20fb^{-1}$ are also plotted under the assumption that the 
background for this decay is negligible.
The corresponding expected sensitivity  can be reach a branching ratio
of $1.3\times 10^{-8}$ and  $6.5\times 10^{-9}$ (dash lines), 
respectively. 
We see that the $R_b$ constraint $f/f'\le 1.9$ 
at 95\% C.L. shown in Fig.~\ref{Fig:bsmumu} is the strongest bound. Comparatively, 
the current upper bound on $Br(B_s\to\mu^+\mu^-)$ from CDF \cite{CDF} is
much weaker, which only excludes a small region with large $f/f'$.
Under the constraint $f/f'\le 1.9$, the enhancement factor for the branching ratio 
in the technicolor model can still be up to 5. The  upgraded Tevatron with  
$20fb^{-1}$ will  be sensitive to enhancements
of \Bsmumu in this model provided that $m_\sigma$ is below
580 GeV.

\section{Conclusions}
\label{conclusion}

We have evaluated the decays \Bsdll in the technicolor model with 
scalars, taking into account various experimental constraints, especially
$R_b$, on the model parameter space. 
We first examined the  restriction on $f/f'$ arising from $R_b$ 
which that previous study did not consider. We found that 
large $f/f'$, which might cause significantly enhancement for 
$Br(B_{s,d}\to\mu^+\mu^-)$ from  neutral scalars in the model, 
has been excluded by the constraints from $R_b$. 
Nevertheless, under the renewed $R_b$ constraint, 
the branching ratio of \Bsmumu  can still be enhanced by a factor of 5 
relative to the SM prediction. With the maximum allowed value of 
$f/f'\sim 1.9$ from $R_b$, 
the upgraded Tevatron with $20fb^{-1}$ will be sensitive to enhancements
of \Bsmumu in this model provided that $m_\sigma$ is below 580 GeV. 
Since the theoretical uncertainties, which primarily come from the B-meson 
decay constants and CKM matrix elements, will be reduced in the on-going 
B-physics experiments and the lattice calculations, the processes \Bsdll 
will promise to be a good probe of new physics.

\end{document}